# Drift-Diffusion Approach to Spin-Polarized Transport

Yuriy V. Pershin

*Center for Quantum Device Technology,*

*Department of Physics and Department of Electrical and Computer Engineering,*

*Clarkson University, Potsdam, New York 13699-5721, USA*

We develop a drift-diffusion equation that describes electron spin polarization density in two-dimensional electron systems. In our approach, superpositions of spin-up and spin-down states are taken into account, this distinguishes our model from the traditional two-component drift-diffusion approximation. The Dresselhaus and Rashba spin-orbit coupling mechanisms are taken into consideration, as well as an applied electric field. The derived equation is applied to the modelling of relaxation of homogeneous spin polarization. Our results are consistent with previous studies.

Recently, considerable progress has been made in research into the fundamental properties of electron spin polarization in semiconductor heterostructures, with the aim to create spin-based electronic devices [1-8]. Different theoretical approaches have been applied to describe the electron spin polarization in semiconductors, including, for example, two-component drift-diffusion model [9-13], the Boltzman equation approach [14-18], a number of Monte Carlo-based techniques [19-26], and microscopic approaches [27-32]. Within the two-component drift-diffusion model, the electrons are considered as particles having spin up or down. The electron spin polarization is defined as $P = n_\uparrow - n_\downarrow$, where $n_{\uparrow(\downarrow)}$ is the density of up-spin (down-spin) electrons, and described by the equation:

$$\frac{\partial P}{\partial t} = D\Delta P + D\frac{e\vec{E}}{k_B T}\nabla P - \frac{P}{\tau} \quad . \tag{1}$$

Here $D$ is the diffusion coefficient, $-e$ is the electron charge, $\vec{E}$ is the electric field, $k_B$ is the Boltzman constant, $T$ is the temperature, and $\tau$ is the spin relaxation time. The two-component drift-diffusion model, based on Eq. (1), was successful in describing electric field effects on spin polarization [9-12]. However, this model is quite limited, since it takes into account only classical states without allowing superposition possible for quantum mechanical states. Superpositions of spin-up and spin-down states, as well as pure and mixed quantum mechanical states, are taken into account in the density matrix formalism [33]. Parameterization of the density matrix by the spin polarizations vector component is very convenient in different applications of this method [21-26,34].

The aim of this paper is to derive a drift-diffusion equation for the spin polarization vector $\vec{P}$ defined as $\vec{P} = Tr(\rho\vec{\sigma})$, where $\rho$ is the single-electron density matrix [33], of a nondegenerate two-dimensional electron gas, which consistently takes into account electric field effects and effects of the spin-orbit interaction. In our calculations we will use a well-known correspondence between quantum mechanical evolution of the spin polarization vector and precession of a classical magnetic moment in an external field. Moreover, we assume that the effect of the spin-orbit interaction on electron space motion is negligible. The main result of our paper, Eq. (7) and Eq. (8), offers a simple approach to calculations of the spin polarization distribution in different systems. At the end of the paper, relaxation of the homogeneous spin polarization is studied.

Let us consider the change of the electron spin polarization $\vec{P}$, in a small two-dimensional volume the size of $\sim L_p^2$, where $L_p$ is the mean free path, near a point $(x_i, y_i)$ (see Fig. 1). Within the nearest-neighbour approximation, where each interior point in the lattice shown in Fig. 1 is bounded orthogonally by four others, the change of the spin polarization in such a volume is considered to be due to the electron diffusion and drift processes through the volume boundaries with four nearest neighbour volumes. In such a lattice, the momentum relaxation time $\tau_p$ defines the hopping time between the lattice points. Assuming that the electric field is in $+x$ direction, the differential equation for $\vec{P}(x_i, y_j)$ is given by

$$\frac{\partial \vec{P}(x_i, y_j)}{\partial t} = \frac{\vec{P}'(x_{i-1}, y_j) + \vec{P}'(x_{i+1}, y_j) + \vec{P}'(x_i, y_{j-1}) + \vec{P}'(x_i, y_{j+1})}{4\tau_p} - \frac{\vec{P}(x_i, y_j)}{\tau_p} + C\left(\vec{P}'(x_{i-1}, y_j) - \vec{P}(x_i, y_j)\right). \quad (2)$$

The first term on the right-hand side of Eq. (2) describes the incoming spin polarization from the neighbor lattice cells, the second term describes the outgoing spin polarization and the last term governs the applied electric field. The coefficient $C$ is proportional to the applied electric field; an explicit expression will be given below. The main difference between the usual rate equation (for example, for a particle density or charge density) and Eq. (2) is hidden in the symbol ', which is used to underline that the incoming spin polarization is changed during the transition process due to the spin-orbit interaction. Such a change of the spin polarization leads to the D'yakonov-Perel' spin relaxation.

There are two main types of spin-orbit interaction in semiconductor heterostructures. The Dresselhaus spin-orbit interaction [35] appears as a result of the asymmetry present in certain crystal lattices, e.g., zinc blende structures. The Rashba spin-orbit interaction [36,37] arises due to the asymmetry associated with the confinement potential and is of interest because of the ability to electrically control the strength of this interaction. The latter is utilized, for instance, in the Datta-Das spin transistor [6]. The Hamiltonian for the Rashba interaction is written [36,37] as

$$H_R = \alpha \hbar^{-1} \left( \sigma_x p_y - \sigma_y p_x \right), \quad (3)$$

where $\alpha$ is the interaction constant, $\vec{\sigma}$ is the Pauli-matrix vector corresponding to the electron spin, and $\vec{p}$ is the momentum of the electron confined in a two-dimensional geometry. For two-dimensional heterostructures with appropriate growth geometry, the Dresselhaus spin-orbit interaction is of the form [35]

$$H_D = \beta\hbar^{-1}\left(\sigma_x p_x - \sigma_y p_y\right), \tag{4}$$

where $\beta$ is the interaction constant. From the point of view of the electron spin, the effect of the spin-orbit coupling can be regarded as an effective magnetic field

$$\vec{H} = \mu_B^{-1}\hbar^{-1}\left(\alpha p_y + \beta p_x,\ -\alpha p_x - \beta p_y\right) \tag{5}$$

acting on the electron spin. Here $\mu_B$ is the Bohr magneton. In the presence of a magnetic field, the electron spins feel a torque and precess in the plane perpendicular to the magnetic field with angular frequency $\vec{\Omega} = \gamma\vec{H}$, where $\gamma$ is the gyromagnetic ratio. The quantum mechanical evolution of the electron spin polarization $\vec{P}$ vector is described by the equation of motion $d\vec{P}/dt = \vec{P}\times\vec{\Omega}$. The angle of the spin rotation per mean free path, $L_P$, is given by $\varphi = 2m^* L_P \sqrt{\alpha^2 + \beta^2}\hbar^{-2}$, where $m^*$ is the effective electron mass. It is important that $\varphi$ does not depend on the electron momentum.

Consider the evolution of the spin polarization during the transfer between two neighbouring lattice cells. Electron spin polarization vector precesses around the effective magnetic field defined by Eq. (5). If $\vec{a}$ is the unit vector along the precession axis, then

$$\vec{P}' = \vec{P} + \vec{P}_\perp(\cos\varphi - 1) + \vec{a}\times\vec{P}\sin\varphi \tag{6}$$

where $\vec{P}_\perp = \vec{P} - \vec{a}(\vec{a}\vec{P})$ is the component of the spin polarization perpendicular to the precession axis. The direction of the precession axis $\vec{a}$ is different for different neighbouring lattice cells. Table 1 gives the direction of the precession axis for different transitions. Substitution of Eq. (6) and $\vec{a}$ from Table 1 into Eq. (2) and application of the continuum limit gives the desired equation for the spin polarization:

$$\frac{\partial \vec{P}}{\partial t} = D\Delta\vec{P} + \frac{(\cos\varphi - 1)}{\tau_p}\left(L_p^2 \Delta\vec{P} + 4\vec{P} - \vec{a}_1\left(\vec{a}_1\left(L_p^2 \frac{\partial^2}{\partial x^2}\vec{P} + 2\vec{P}\right)\right) - \vec{a}_1\left(\vec{a}_1\left(L_p^2 \frac{\partial^2}{\partial y^2}\vec{P} + 2\vec{P}\right)\right)\right) - 2L_p \frac{\sin\varphi}{\tau_p}\left(\vec{a}_1 \times \frac{\partial \vec{P}}{\partial x} + \vec{a}_2 \times \frac{\partial \vec{P}}{\partial y}\right) +$$
$$C\left(-L_p \frac{\partial \vec{P}}{\partial x} + (\cos\varphi - 1)\left(\vec{P} - L_p \frac{\partial \vec{P}}{\partial x} - \vec{a}_1\left(\vec{a}_1\left(\vec{P} - L_p \frac{\partial \vec{P}}{\partial x}\right)\right)\right) + \sin\varphi\, \vec{a}_1 \times \left(\vec{P} - L_p \frac{\partial \vec{P}}{\partial x}\right)\right), \tag{7}$$

with $D = L_p^2/4\tau_p$. Comparing the leading term of Eq. (7) with Eq. (1) we find $C = -DeE/(k_B T L_p)$. Assuming that the elementary rotations during free flight time are small and $\vec{P}$ is a smooth slowly-varying function of $(x,y)$, Eq. (7) can be rewritten as

$$\frac{\partial \vec{P}}{\partial t} = D\Delta\vec{P} - CL_p\frac{\partial \vec{P}}{\partial x} - \frac{2L_p\varphi}{\tau_p}\left(\vec{a}_1 \times \frac{\partial \vec{P}}{\partial x} + \vec{a}_2 \times \frac{\partial \vec{P}}{\partial y}\right) - \frac{\varphi^2}{\tau_p}\left(2\vec{P} - \vec{a}_1(\vec{a}_1\vec{P}) - \vec{a}_2(\vec{a}_2\vec{P})\right) + C\left(-\frac{\varphi^2}{2}(\vec{P} - \vec{a}_1(\vec{a}_1\vec{P})) + \varphi\vec{a}_1 \times \left(\vec{P} - L_p\frac{\partial \vec{P}}{\partial x}\right)\right)$$

(8)

We shall now highlight the meaning of all the terms in the right-hand side of Eq. (8). The first two terms describe the diffusion of the spin polarization in the electric field without relaxation, they are similar to the first two terms in the right-hand side of Eq. (1). The third term is responsible for the evolution of non-homogeneous spin polarization. The forth term is responsible for the D'yakonov-Perel' relaxation [38,39], and the last term describes the spin polarization precession associated with the applied electric field. It should be noted that an equation, that closely resembles Eq. (8), has been recently derived by using the Wigner function formalism in Ref. [30]. However, Eq. (7) of the present paper is more general, because it was derived for arbitrary spin precession angle per mean free path and contains some additional terms that allow a more precise description of spin polarization dynamics.

Our model allows sufficiently simple and not time-consumingmodeling of various effects related to the spin polarization evolution and distribution in semiconductor heterostructures, as compared with Monte-Carlo techniques [19-26]. Modeling of a particular physical system requires supplemental initial and boundary conditions for Eq. (8). Sources of spin polarization can be introduced into calculations via specifying the magnitude of the spin polarization at the boundary [9,10], or via introduction of additional terms in the right-hand side of Eq. (8) describing the spin injection processes [11]. For example, an instantaneous source of spin polarization can be described by the term $F_0 f(x,y)\delta(t)$, where $F_0$ measures the spin polarization density created at the initial moment of time and the function $f(x,y)$ gives the shape of initial spin polarization and normalized to 1. The continuous point source of spin polarization is described $G_0\delta(x-x_0, y-y_0)$, where $G_0$ measures the spin polarization density created per unit time. When dealing with systems having restrictions on electron space motion, the flux of the spin polarization through the boundary should be set equal to zero. It is worth remembering that this approach assumes that the system sizes are much larger than the electron mean free path. Solutions to the linearized equation (8) in some particular cases are presented below.

Let us consider evolution of homogeneous spin polarization of two-dimensional infinite electron gas in zero applied electric field. In this case Eq. (8) reduces to

$$\frac{\partial \vec{P}}{\partial t} = -\frac{\varphi^2}{\tau_p}\left(2\vec{P} - \vec{a}_1(\vec{a}_1\vec{P}) - \vec{a}_2(\vec{a}_2\vec{P})\right) \qquad (9)$$

Equations for spin polarization components have the form:

$$\frac{\partial P_x}{\partial t} = -\frac{\varphi^2}{\tau_p}\left(P_x + \frac{2\alpha\beta}{\alpha^2+\beta^2}P_y\right)$$

$$\frac{\partial P_y}{\partial t} = -\frac{\varphi^2}{\tau_p}\left(P_y + \frac{2\alpha\beta}{\alpha^2+\beta^2}P_x\right) \qquad (10)$$

$$\frac{\partial P_z}{\partial t} = -\frac{2\varphi^2}{\tau_p} P_z$$

Solutions of Eqs. (10) are given by

$$P_{x,y} = \frac{\left(P_x^0 + P_y^0\right)}{2} e^{-\frac{\varphi^2}{\tau_p}\left(1 + \frac{2\alpha\beta}{\alpha^2 + \beta^2}\right)t} \pm \frac{\left(P_x^0 - P_y^0\right)}{2} e^{-\frac{\varphi^2}{\tau_p}\left(1 - \frac{2\alpha\beta}{\alpha^2 + \beta^2}\right)t} \tag{11}$$

and

$$P_z(t) = P_z^0 e^{-\frac{2\varphi^2 t}{\tau_p}} \tag{12}$$

Here $\left(P_x^0, P_y^0, P_z^0\right)$ is the initial spin polarization. The important feature of the solutions (11), (12) is that the evolution of the spin polarization is governed by three relaxation times. These times describe evolution of different components of the spin polarization, specifically, projections of the spin polarization on $(1,1,0)$, $(1,-1,0)$ directions and $z$ component of the spin polarization. The corresponding relaxation times are: $\tau_2 = \tau_p / \left(K(\alpha+\beta)^2\right)$, $\tau_3 = \tau_p / \left(K(\alpha-\beta)^2\right)$, and $\tau_1 = \tau_p / \left(2K(\alpha^2 + \beta^2)\right)$, where $K = \left(2m^* L_P \hbar^{-2}\right)^2$. These results are in full agreement with Ref. [40]. The relaxation time dependences on the Rashba spin-orbit coupling constant $\alpha$ at a fixed value of the Dresselhaus spin-orbit coupling constant $\beta$ are shown in Fig. 2. When $\alpha = 0$, the in-plane relaxation times are two times longer than the relaxation time for $z$ component of the spin polarization, in correspondence with Ref. [38,39]. The relaxation time $\tau_3$ dramatically changes with the change of $\alpha$: when $\alpha = \beta$ this time is infinity. This remarkable behavior of the spin relaxation time is used in the recent proposal of a spin-transistor [41]. However, relaxation of spin polarization projection on $(1,-1,0)$ direction is still possible via alternative spin relaxation mechanisms [34,42-46].

Another interesting case under consideration is the relaxation of the homogeneous spin polarization in a quantum wire in zero applied electric field. The equation for spin polarization in 1D has a form

$$\frac{\partial \vec{P}}{\partial t} = -\frac{\varphi^2}{\tau_p}\left(\vec{P} - \vec{a}_1\left(\vec{a}_1 \vec{P}\right)\right) \tag{13}$$

Solution of Eq. (13) is given by $\vec{P}(t) = \vec{P}_\| + \vec{P}_\perp \exp\left(-\varphi^2 t / \tau_p\right)$, where $\vec{P}_\|$ is the component of the spin polarization parallel to $\vec{a}_1$. It should be noted that this solution is valid with the understanding that the transverse electron confinement is strong and a single subband corresponding to the transverse confinement is occupied. If transverse confinement is weak, several transverse subbands could be occupied and mechanism of spin relaxation due to the different spin-orbit coupling constants in different subbands becomes operative [22]. It leads to relaxation of $\vec{P}_\|$ as well.

In summary, this paper has focused on derivation of drift-diffusion equation for spin polarization in two-dimensional electron gas with spin-orbit interaction in an applied electric field. We have derived such an equation using a lattice model for spin polarization diffusion accounting for precession of spin polarization vector in an effective magnetic field. The obtained equation has been solved in assumption of homogeneous spin polarization of 1D and 2D electron gas. Our results show agreement with previous studies.

We gratefully acknowledge helpful discussions with Prof. V. Privman. This research was supported by the National Security Agency and Advanced Research and Development Activity under Army Research Office contract DAAD-19-02-1-0035, and by the National Science Foundation, grant DMR-0121146.

**Table 1.** Electron spin precession axis for the incoming spin polarization to the cell $(x_i, y_j)$ from neighbour cells.

| Cell | $(x_{i-1}, y_j)$ | $(x_{i+1}, y_j)$ | $(x_i, y_{j-1})$ | $(x_i, y_{j+1})$ |
|---|---|---|---|---|
| $\vec{a}$ | $\vec{a}_1 = (\beta, -\alpha)/\sqrt{\beta^2 + \alpha^2}$ | $-\vec{a}_1$ | $\vec{a}_2 = (\alpha, -\beta)/\sqrt{\beta^2 + \alpha^2}$ | $-\vec{a}_2$ |

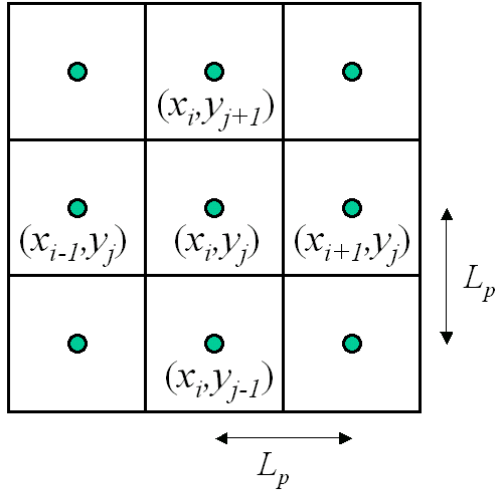

FIG. 1. Lattice grid.

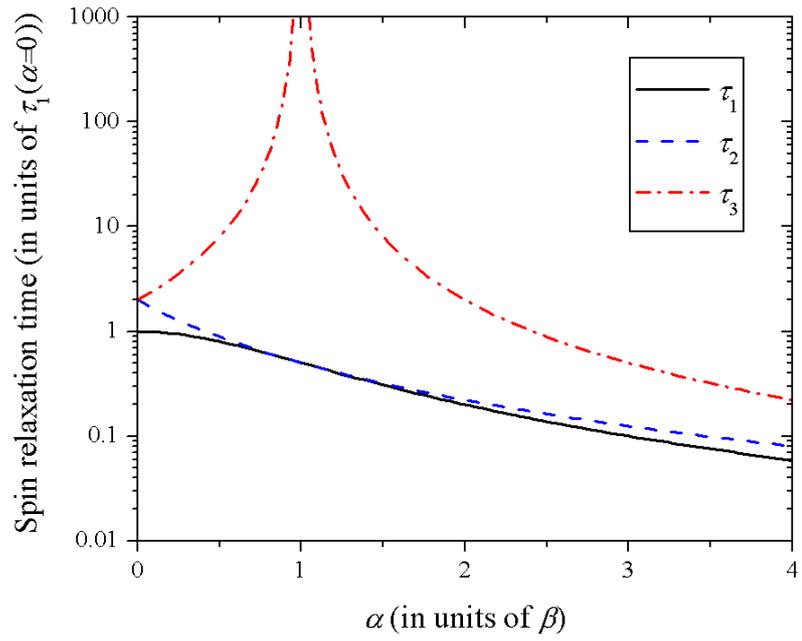

FIG.2. Spin relaxation times as a function of Rashba spin-orbit coupling constant $\alpha$.

10